\documentclass{article}
\usepackage[T1]{fontenc}
\usepackage{wrapfig}
\usepackage[portuges,english]{babel}
\usepackage{amssymb,latexsym,amsmath,color,mathrsfs,ifsym,graphics,stmaryrd} 
\usepackage[colorlinks,linkcolor=blue,urlcolor=blue,citecolor=black,
plainpages=false,pdfpagelabels,breaklinks]{hyperref}

\title{{\sc Measuring Quantum Superpositions\\ 
\smallskip
\smallskip
(Or, ``It is only the theory which\\ decides what can be observed.'')}}

\author{{\sc Christian de Ronde}\thanks{Fellow Independent Researcher of the Consejo
Nacional de Investigaciones Cient\'{\i}ficas y T\'ecnicas.}}
\date{}

\usepackage[margin=2 cm]{geometry}

\begin{document}
\maketitle

\begin{center}
\begin{small}
Philosophy Institute ``Dr. A. Korn'' Buenos Aires University, CONICET \\
Engineering Institute, National University Arturo Jauretche - Argentina. \\
Federal University of Santa Catarina - Brasil. \\
Center Leo Apostel fot Interdisciplinary Studies, Brussels Free University - Belgium. \\
\end{small}
\end{center}
\begin{abstract}
\noindent In this work we attempt to confront the orthodox widespread claim present in the foundational literature of Quantum Mechanics (QM) according to which `superpositions are never actually observed in the lab'. In order to do so, we begin by providing a critical analysis of the famous measurement problem which, we will argue, was originated by the strict application of the empirical-positivist requirements to subsume the quantum formalism under their specific understanding of `theory'. In this context, the {\it ad hoc} introduction of the {\it projection postulate} (or measurement rule) can be understood as a necessary requirement coming from a naive empiricist standpoint which presupposes that {\it observations} are self evident {\it givens} of ``common sense'' experience ---independent of metaphysical (categorical) presuppositions. We then turn our attention to two ``non-collapse'' interpretations of QM ---namely, modal and many worlds--- which even though deny that the ``collapse'' is a real physical process anyhow retain the measurement rule as a necessary element of the theory. In contraposition, following Einstein's claim according to which ``it is only the theory which decides what can be observed'', we propose a return to the realist representational understanding of  `physical theories' in which `observation' is considered as derived from theoretical presuppositions. It is from this standpoint that we discuss a new non-classical conceptual representation which allows us to understand quantum phenomena in an intuitive ({\it anschaulicht}) manner. Leaving behind the projection postulate, we discuss the general physical conditions for measuring and observing quantum superpositions. 
\end{abstract}

\renewenvironment{enumerate}{\begin{list}{}{\rm \labelwidth 0mm
\leftmargin 0mm}} {\end{list}}

\newcommand{\ita}{\textit}
\newcommand{\mcal}{\mathcal}
\newcommand{\mfrak}{\mathfrak}
\newcommand{\mbb}{\mathbb}
\newcommand{\mrm}{\mathrm}
\newcommand{\msf}{\mathsf}
\newcommand{\mscr}{\mathscr}
\newcommand{\lra}{\leftrightarrow}
\renewenvironment{enumerate}{\begin{list}{}{\rm \labelwidth 0mm
\leftmargin 5mm}} {\end{list}}

\newtheorem{dfn}{\sc{Definition}}[section]
\newtheorem{thm}{\sc{Theorem}}[section]
\newtheorem{lem}{\sc{Lemma}}[section]
\newtheorem{cor}[thm]{\sc{Corollary}}
\newcommand{\Proof}{\textit{Proof:} \,}
\newcommand{\cqd}{{\rule{.70ex}{2ex}} \medskip}

\bigskip

\bigskip

\bigskip

\bigskip

\bigskip

\begin{flushright}
{\it The highest would be: to realize that\\ everything factual is already theory.\\}
\smallskip
Goethe. 
\end{flushright}
\section*{Introduction}

Within the philosophical and foundational literature about Quantum Mechanics (QM) there exists a widespread idea according to which superposed states are never observed in the lab. The orthodox claim is that we never find in our macroscopic world, after a quantum measurement has been performed, the pointer of an apparatus in a superposed state. Instead, it is argued that what we actually observe is a single outcome; i.e., a `click' in a detector or a `spot' in a photographic plate. Since according to the orthodox empirical-positivist understanding, theories must be able to account for what we actually observe, it appears to be something really wrong with QM: the theory simply does not seem to provide an account of the empirical observations it should be talking about. In order to fill this void, during the axiomatization of the theory which took place at the beginning of the 1930s, Paul Dirac and John von Neumann introduced as an axiom of the theory itself a rule that would allow them to turn any quantum superposition into only one of its terms ---representing the observed outcome. The addition of this axiom had as a consequence the introduction of a new invisible process that was not described by the theory and went explicitly against the linearity of the whole mathematical formalism. Suddenly, there were two different evolutions within QM: firstly, a deterministic evolution described by Schr\"odinger's equation of motion which worked perfectly well when no one was looking; and secondly, a strange indeterministic evolution ---a ``collapse'' of the superposition to only one of its terms--- produced each time a measurement was actually performed. Very soon, the Dirac-von Neumann axiomatic presentation, in tune with the positivist {\it Zeitgeist} of the epoch, became standardized as the orthodox textbook formulation of QM. Strange as it might seem, while collapses became accepted as part of the theory, quantum superpositions ---an essential part of the mathematical formalism and the main reason for introducing collapses--- became to be regarded with great skepticism and even despise (see \cite{deRonde18} for a detailed analysis). One of the main reasons behind this general attitude present within the orthodox understanding of QM might be found in the difficulties of providing a conceptual representation that would match the mathematical features implied by superposed states. Dirac \cite[p. 12]{Dirac74} had already realized these difficulties pointing to the fact that: ``The nature of the relationships which the superposition principle requires to exist between the states of any system is of a kind that cannot be explained in terms of familiar physical concepts.'' This became even more explicit after the famous {\it Gedankenexperiment} proposed by Erwin Schr\"odinger in 1935 which ---following the ideas of Einstein--- exposed the fact that quantum superpositions could not be understood in terms of classical entities ---constituted categorically in terms of the principles of existence, non-contradiction and identity. Schr\"odinger used a `cat' in what could be considered an {\it ad absurdum} proof of the impossibility to represent (mathematical) quantum superpositions in terms of the classical notion of object; i.e., a system with non-contradictory definite valued properties. Not only did he demonstrate that the properties of being `alive' and `dead' were quantified beyond the {\it certainty} of binary valuations, he also showed that, since an ``ignorance interpretation'' of the sates was precluded by the mathematical formalism itself, these contradicting properties had to be both considered as somehow truly existent. Unfortunately, the influence of the Bohrian-positivist alliance converging after the war in the instrumentalist credence of physicists, precluded the possibilities of a deeper analysis. It was only when Alain Aspect's famous experiment during the first years of the 1980s ---testing Boole-Bell inequalities in an EPR situation--- showed that both quantum superpositions and entanglement ---as discussed and critically analyzed by both Einstein and Schr\"odinger--- could be used as a resource for information processing that things became to really change. As described by Jeffrey Bub \cite{Bub17}, ``[...] it was not until the 1980s that physicists, computer scientists, and cryptographers began to regard the non-local correlations of entangled quantum states as a new kind of non-classical resource that could be exploited, rather than an embarrassment to be explained away.''  As he continues to explain: ``Most physicists attributed the puzzling features of entangled quantum states to Einstein's inappropriate `detached observer' view of physical theory, and regarded Bohr's reply to the EPR argument (Bohr, 1935) as vindicating the Copenhagen interpretation. This was unfortunate, because the study of entanglement was ignored for thirty years until John Bell's reconsideration of the EPR argument (Bell, 1964).'' In this new context of technical possibilities, the ``shut up and calculate!'' widespread instrumentalist attitude of post-war physicists had to unwillingly allow the reopening of foundational and philosophical debates about QM ---which had been almost completely silenced in the literature for almost half a century. However, instead of focusing on the conceptualization of superpositions and entanglement, it was the justification of the measurement rule, known in the literature as `the measurement problem', which became placed at the center of the philosophical stage. The fact that in the time being no one had ever been able to experimentally test in the lab the actual existence of such ``collapses'' did not seem very important for a physics community that had been trained learning that QM could not be understood.\footnote{As Lee Smolin \cite[p. 312]{Smolin07} would describe the post-war instrumentalist view of physics recalling his own experience as a student: ``When I learned physics in the 1970s, it was almost as if we were being taught to look down on people who thought about foundational problems. When we asked about the foundational issues in quantum theory, we were told that no one fully understood them but that concern with them was no longer part of science. The job was to take quantum mechanics as given and apply it to new problems. The spirit was pragmatic; `Shut up and calculate' was the mantra. People who couldn't let go of their misgivings over the meaning of quantum theory were regarded as losers who couldn't do the work.''} The obvious conclusion that the collapse was an artificial fiction which led nowhere was never seriously considered and Professors kept teaching young students in Universities all around the world about the existence of a measurement rule within the theory. 

Confronting orthodoxy, in this work we attempt to provide a critical analysis of the general conditions for observing and measuring quantum superpositions. We begin, in section 1, by analyzing the essential link between the {\it ad hoc} introduction of the projection postulate and the positivist reference to ``common sense'' observations in physical theories. In section 2 we discuss the role of ``collapses'' in different contemporary interpretations of QM. In particular, we address two ``non-collapse'' proposals, namely, the many worlds and the modal interpretations. Section 3 discusses the missing link between the orthodox mathematical formalism of QM and its conceptual framework. In section 4, we revisit what is actually observed in the lab and in section 5, we show how an objective conceptual account of QM can be derived from the operational-invariance present within the mathematical formalism of the theory itself. From this standpoint, in section 5, we discuss the general conditions for measuring and observing quantum superpositions in the lab.  
 


\section{Positivism, Quantum Theory and the Measurement Rule}

Positivist ideas had a completely different impact in QM during the creation of the theory, and then, during its development and progressive standardization. In the second half of 19th Century positivism played a subversive role fighting against the dogmatism imposed by classical Newtonian physics and its {\it a priori} Kantian metaphysical interpretation. During the early 20th Century positivism played an important guiding role helping the new generation of physicists to escape the metaphysical presuppositions imposed by classical physics. The result of this adventure beyond the limits of the classical representation was the creation of relativity theory and QM. However, once QM had been finally developed by Werner Heisenberg as a closed mathematical formalism in 1925, positivism begun to play an oppressive role, limiting the possibilities of development and understanding of the same theory it had ---undoubtedly--- helped to create. It is this latter approach which David Deutsch \cite[p. 308]{Deutsch04} ---pointing explicitly to empiricism, positivism, Bohr and instrumentalism\footnote{As Deutsch \cite[p. 312]{Deutsch04} remarks: ``[...] empiricism did begin to be taken literally, and so began to have increasingly harmful effects. For instance, the doctrine of positivism, developed during the nineteenth century, tried to eliminate from scientific theories everything that had not been `derived from observation'. Now, since nothing is ever derived from observation, what the positivists tried to eliminate depended entirely on their own whims and intuitions.'' Regarding the Danish physicist, Deutsch \cite[p. 308]{Deutsch04} makes the point that:  ``The physicist Niels Bohr (another of the pioneers of quantum theory) then developed an `interpretation' of the theory which later became known as the `Copenhagen interpretation'. It said that quantum theory, including the rule of thumb, was a complete description of reality. Bohr excused the various contradictions and gaps by using a combination of instrumentalism and studied ambiguity. He denied the `possibility of speaking of phenomena as existing objectively'  ---but said that only the outcomes of observations should count as phenomena. He also said that, although observation has no access to `the real essence of phenomena', it does reveal relationships between them, and that, in addition, quantum theory blurs the distinction between observer and observed. As for what would happen if one observer performed a quantum-level observation on another, he avoided the issue.''}--- has characterized as `bad philosophy', namely, ``[a] philosophy that is not merely false, but actively prevents the growth of other knowledge.'' Indeed, during the subsequent years, Niels Bohr developed an interpretation which embraced the impossibility of a consistent theoretical representation, something that he became to regard as the very essence of the quantum revolution. In this respect, his principles of {\it complementarity} and {\it correspondence}, together with fictional images like `quantum particles', `quantum waves' and `quantum jumps', allowed him to support a pragmatic understanding of the theory of quanta ---also close to the positivist principles. Finally, in 1930, a young English engineer and mathematician named Paul Dirac was able to bring together both Bohrian and positivist ideas in his book, {\it The Principles of Quantum Mechanics} \cite{Dirac74}. Dirac presented an axiomatic vectorial formulation of QM which attempted to unify the earlier developments, firstly by Werner Heisenberg, Pascual Jordan and Max Born in 1925, and one year later by Erwin Schr\"odinger. Following both positivism and Bohr's of account of QM, Dirac begun by stressing that ``[it is] important to remember that science is concerned only with observable things and that we can observe an object only by letting it interact with some outside influence. An act of observation is thus necessarily accompanied by some disturbance of the object observed.'' Continuing with Bohr's interpretation of Heisenberg's inequalities, Dirac remarked that: ``we have to assume that {\it there is a limit to the finiteness of our powers of observation and the smallness of the accompanying disturbance ---a limit which is inherent in the nature of things and can never be surpassed by improved technique or increased skill on the part of the observer.}'' Adopting the positivist standpoint Dirac claimed that in science we gain knowledge only through observation, but embracing Bohr's ideas we must also accept that in QM we have reached a limit to the possibility of observability and representation itself. Dirac's scheme was then confronted to the mathematical formalism and the existence of quantum superpositions which he himself had recognized as the kernel constituents of the theory. 
\begin{quotation}
\noindent {\small``The nature of the relationships which the superposition principle requires to exist between the states of any system is of a kind that cannot be explained in terms of familiar physical concepts. One cannot in the classical sense picture a system being partly in each of two states and see the equivalence of this to the system being completely in some other state. There is an entirely new idea involved, to which one must get accustomed and in terms of which one must proceed to build up an exact mathematical theory, without having any detailed classical picture.'' \cite[p.12]{Dirac74}}\end{quotation}
The fact that superpositions did not provide by themselves a consistent representation of what was going on did not seem like a problem. After all, Bohr had not provided a consistent representation of his model of the atom nor the existence of quantum jumps. As Dirac would famously argue:  ``it might be remarked that the main object of physical science is not the provision of pictures, but the formulation of laws governing phenomena and the application of these laws to the discovery of phenomena. If a picture exists, so much the better; but whether a picture exists of not is a matter of only secondary importance.'' What remained a problem for Dirac's positivist standpoint was the relation between these weird mathematical elements and the need to predict observability with {\it certainty}; i.e., as a {\it yes-no binary reply} characterizing the result of a given measurement situation. According to the positivist understanding, a formalism ``becomes a precise physical theory when all the axioms and rules of manipulation governing the mathematical quantities are specified and when in addition certain laws are laid down connecting physical facts with the mathematical formalism''. It is at this point that, making use of an example of polarized photons, Dirac introduced today's famous ``collapse'' of the quantum wave function: 
\begin{quotation}
\noindent {\small``When we make the photon meet a tourmaline crystal, we are subjecting it to an observation. We are observing wither it is polarized parallel or perpendicular to the optic axis. The effect of making this observation is to force the photon entirely into the state of parallel or entirely into the state of perpendicular polarization. It has to make a sudden jump from being partly in each of these two states to being entirely in one or the other of them. Which of the two states it will jump cannot be predicted, but is governed only by probability laws.'' \cite[p. 9]{Dirac74}} 
\end{quotation}
Two years later, in 1932, a Hungarian mathematician called John von Neumann turned the ``jump'' into a measurement postulate of the theory itself, the famous {\it Projection Postulate} \cite{VN}. The books by Dirac and von Neumann were considered by mathematical physicists as a sound and rigorous formulation of the theory and very soon became the orthodox understanding of QM. It is in this way that both projection postulate and the existence of ``collapses'' became accepted within the standard textbook formulation of the theory of quanta. However, from a realist viewpoint, the measurement rule and the collapses were clearly problematic ---to say the least. It is in this context that the so called ``measurement problem'' became part of the main philosophical debate surrounding QM:

\smallskip 
\smallskip 

\noindent {\it {\bf Quantum Measurement Problem (QMP):} Given a specific basis (or context), QM describes mathematically a quantum state in terms of a superposition of, in general, multiple states. Since QM allows us to predict that the quantum system will get entangled with the apparatus and thus its pointer positions will also become a superposition,\footnote{Given a quantum system represented by a superposition of more than one term, $\sum c_i | \alpha_i \rangle$, when in contact with an apparatus ready to measure, $|R_0 \rangle$, QM predicts that system and apparatus will become ``entangled'' in such a way that the final `system + apparatus' will be described by  $\sum c_i | \alpha_i \rangle  |R_i \rangle$. Thus, as a consequence of the quantum evolution, the pointers have also become ---like the original quantum system--- a superposition of pointers $\sum c_i |R_i \rangle$. This is why the measurement problem can be stated as a problem only in the case the original quantum state is described by a superposition of more than one term.} the question is why do we observe a single outcome instead of a superposition of them?}

\smallskip 
\smallskip 

In order to understand the centrality of the measurement problem in QM one needs to pay special  attention to the 20th Century positivist re-foundation of physics which imposed within science a radically new meaning and understanding of physical theories ---following Ernst Mach--- as ``economies of experience''. This radical shift took place through the introduction of a new scheme grounded on two main pillars: on the one hand, ``common sense'' observations, from which theories were derived; and on the other, mathematical formalisms which, in turn, allowed to predict future observations. Physical theories were not understood anymore as providing a formal-conceptual representations of a state of affairs but instead, as mathematical apparatuses capable of predicting ---through a series of rules--- observations in the lab.\footnote{Technically speaking, the distinction between {\it empirical terms} (i.e., the empirically ``given'') and {\it theoretical terms} (i.e., their translation into simple statements) comprised this new understanding of theories.} In fact, the measurement problem was a consequence of the imposed need to fulfill the positivist requirement of predicting single measurement outcomes ---something which Heisenberg's matrix formulation clearly made no reference to. 

It is also important to stress that both positivist philosophers and Bohr, as the main leader of the physics community, were very careful not to loose all contact with conceptual (or metaphysical) representation. In fact, as some kind of metaphysical ghost that would refuse to completely vanish, the reference to (microscopic) `particles' would remain always present in the background of their discourse. Furthermore, it became accepted that once the mathematics of an empirically adequate theory was constructed, it was also possible ---for those willing to do so--- to add an `interpretation' that would explain how the world was according to the theory. In this way, anti-realist positivism was able to include both empiricism and realism within its own scheme. But of course, since physical phenomena was now understood as a ``self-evident'' {\it given} ---independent of physical concepts and metaphysical presuppositions---, the introduction of an `interpretation' (of the mathematical formalism) was not really necessary. Any empirically adequate theory already did its job as a predictive device without the addition of a metaphysical narrative.\footnote{It is important to remark that the meaning of `metaphysics' in the positivist context was understood as an unjustified discourse about the un-observable.} This obviously implied that in the positivist scheme, empiricism was much more fundamental than the degraded form of (metaphysical) realism it also supported. In a very astute manner empirical positivism had followed the recommendation by Michael Corleone in {\it The Godfather}: ``Keep your friends close, and your enemies closer.'' In order to retain a realist discourse, it was accepted by the new anti-realist trend of thought that an empirically adequate theory could have many different `interpretations'. Many different narratives could be added in order to explain what a theory was really talking about beyond {\it hic et nunc} observations. Thus, realism became a kind of ``dressing'' for theories,  something to be added by metaphysically inclined physicists and philosophers which continued to stubbornly wonder about a reality beyond the observable realm ---something that became to be known as ``external reality'' in contraposition to the ``internal (or empirical) reality'' of the subject. As remarked by van Fraassen \cite[pp. 202-203]{VF80}: ``To develop an empiricist account of science is to depict it as involving a search for truth only about the empirical world, about what is actual and observable.'' But, of course, regarding the reference of the theory \cite[p. 242]{VF91} beyond observations: ``However we may answer these questions [regarding the interpretation], believing in the theory being true or false is something of a different level.'' After centuries of confrontations, anti-realism had finally been able to confine realism into a small prison cell. Presenting it as a mere fictional narrative with no proper fundament, anti-realism was able to diminish the fundamental role of reality (or {\it physis}) within physics. As Karl Popper would famously remark:  
\begin{quotation}
\noindent {\small ``The empirical basis of objective science has thus nothing `absolute' about it. Science does not rest upon solid bedrock. The bold structure of its theories rises, as it were, above a swamp. It is like a building erected on piles. The piles are driven down from above into the swamp, but not down to any natural or `given' base; and if we stop driving the piles deeper, it is not because we have reached firm ground. We simply stop when we are satisfied that the piles are firm enough to carry the structure, at least for the time being.'' \cite[p. 111]{Popper92}} 
\end{quotation}
Positivism portrayed realism as a subjective belief in `pictures' and `images' created by people who preferred ungrounded metaphysical bla bla instead of down to earth perception. As David Deutsch makes the point:   
\begin{quotation}
\noindent {\small ``During the twentieth century, anti-realism became almost universal among philosophers, and common among scientists. Some denied that the physical world exists at all, and most felt obliged to admit that, even if it does, science has no access to it. For example, in {\it Reflections on my Critics} the philosopher Thomas Kuhn wrote: `There is [a step] which many philosophers of science wish to take and which I refuse. They wish, that is, to compare [scientific] theories as representations of nature, as statements about {\it what is really out there}'.'' \cite[p. 313]{Deutsch04}} 
\end{quotation}

However, the main positivist claim according to which {\it empirical terms} (or observations) could be understood without making reference to conceptual (or metaphysical) presuppositions remained completely unjustified. This problem ---which Kant had already discussed two Centuries before in his {\it Critique of Pure Reason}--- was repeatedly addressed by the main figures of both positivism and post-positivism.  Rudolph Carnap \cite{Carnap28}, Ernst Nagel \cite{Nagel61}, Popper \cite{Popper92} and many others tried to present a solution without any success. And even though in the 1950s it was explicitly recognized by Norwood Hanson \cite{Hanson} ---and later on, during the 1960s, by Thomas Kuhn and Paul Feyerabend--- that observation in science could not be considered without a presupposed theoretical scheme (i.e., the {\it theory-ladenness} of observations), the general naive empiricist framework remained completely untouched. More importantly, the problems constructed in both physics and philosophy of physics continued to ---either implicitly or explicitly--- accept naive empiricism as a main standpoint of analysis. The failure of the whole empirical-positivist program which was implicitly recognized in another famous paper by Carl Hempel \cite{Hempel58}, did not change anything. Regardless of the deep internal criticisms and the lack of answers, naive empiricism has continued to play an essential role in the discussions and debates within both physics and philosophy of physics. QM might be regarded as one of the most explicit exposures of the the empirical-positivist influence within the physical sciences. This scheme of understanding of theories is not only responsible for having created the (in)famous measurement problem of QM, it is also guilty for maintaining it ---still today--- at the very center of foundational and philosophical debates.

\section{Measurement Rules, Quantum Collapses and Interpretations}

After the Second World War, the Bohrian-positivist alliance converged into a new pragmatic trend of thought even more radical than its predecessors. Instrumentalism, as part of the 20th Century anti-realist {\it Zeitgeist}, was presented by the the U.S. philosopher John Dewey as the natural extension of both pragmatism and empirical positivism. While pragmatism sustained that the value of an idea is determined by its usefulness, instrumentalism, rejecting the need of any metaphysical fundament, was ready to take a step further and claim that the question regarding the {\it reference} of theories was simply meaningless. The main unspoken claim surrounding the Bohrian-positivist anti-realist alliance was now ready to see the light: {\bf Scientific theories do not make reference to an underlying reality.} Thus, there is no sense in which a theory can be said to be {\it true} or {\it false} (or better or worse) apart from the extent to which it is useful as a ``tool'' in solving scientific problems.\footnote{ ``In the US, which after the Second World War became the central stage of research in physics in the West, the discussions about the interpretation of quantum mechanics had never been very popular. A common academic policy was to gather theoreticians and experimentalists to gather in order to favour experiments and concrete applications, rather than abstract speculations. This practical attitude was further increased by the impressive development of physics between the 1930s and the 1950s, driven on the one hand by the need to apply the new quantum theory to a wide range of atomic and subatomic phenomena, and on the other hand by the pursuit of military goals. As pointed out by Kaiser, `the pedagogical requirements entailed by the sudden exponential growth in graduate student numbers during the cold war reinforced a particular instrumentalist approach to physics'.'' \cite[pp. 2-3]{Osnaghi09}} Popper \cite{Popper63}, a post-positvist himself, concluded at the beginning of the 1960s that instrumentalism had finally conquered the whole of physics: ``Today the view of physical science founded by Osiander, Cardinal Bellarmino, and Bishop Berkeley, has won the battle without another shot being fired. Without any further debate over the philosophical issue, without producing any new argument, the {\it instrumentalist} view (as I shall call it) has become an accepted dogma. It may well now be called the `official view' of physical theory since it is accepted by most of our leading theorists of physics (although neither by Einstein nor by Schr\"odinger). And it has become part of the current teaching of physics.'' As a consequence of the triumph of instrumentalism, QM has been taught ever since as a ``recipe'' to compute measurement outcomes. As recently described by Tim Maudlin: 
\begin{quotation}
\noindent {\small ``What is presented in the average physics textbook, what students learn and researchers use, turns out not to be a precise physical theory at all. It is rather a very effective and accurate recipe for making certain sorts of predictions. What physics students learn is how to use the recipe. For all practical purposes, when designing microchips and predicting the outcomes of experiments, this ability suffices. But if a physics student happens to be unsatisfied with just learning these mathematical techniques for making predictions and asks instead what the theory claims about the physical world, she or he is likely to be met with a canonical response: Shut up and calculate!'' \cite[pp. 2-3]{Maudlin19}} 
\end{quotation} 
The `interpretation' of QM was something left for a minority of physicists who worried about the reference of the theory to physical reality were able to find shelter in philosophy departments. But, maybe as a kind of reaction to the anti-representational ``shut up and calculate!'' instrumentalist program, with the establishemnt of the new field of philosophy of QM ---during the 1980s--- the interpretations of the theory of quanta begun to reproduce themselves at an amazing speed creating what Adan Cabello has recently characterized as ``a map of madness'' \cite{Cabello17}. Without any clear methodology, conceived as metaphysical narratives that would discuss the ontological reference of theories beyond the observable realm, interpretations of QM have expanded in all possible directions. One of the main groups, commanded by Heisenberg, addressed the possibility of returning to an hylomorphic metaphysical scheme in which the realm of potentiality would play an essential role for quantum particles \cite{Heis58}. Henry Margenau, Gilbert Simondon, and even Popper, followed this metaphysical path adding new notions such as latencies, propensities and dispositions (see for a detailed analysis \cite{deRonde17}). However, as remarked by Mauro Dorato with respect to propensity or dispositionalist type interpretations: 
\begin{quotation}
\noindent {\small ``[...] dispositions express, directly or indirectly, those regularities of the world around us that enable us to predict the future. Such a predictive function of dispositions should be attentively kept in mind when we will discuss the `dispositional nature' of microsystems before measurement, in particular when their states is not an eigenstate of the relevant observable. In a word, the use of the language of `dispositions' does not by itself point to a clear ontology underlying the observable phenomena, but, especially when the disposition is irreducible, refers to the predictive regularity that phenomena manifest. {\it Consequently, attributing physical systems irreducible dispositions, even if one were realist about them, may just result in more or less covert instrumentalism.}'' \cite[p. 4]{Dorato06} (emphasis added)}
\end{quotation}
As we have remarked in \cite{deRonde17}, Dorato's criticism to dispositions can be easily extended to all hylomorphic approaches which have failed to provide a clear account of their metaphysical content. Instead of trying to provide a conceptual reference to the mathematical formalism, another important group of interpretations have followed the opposite path. Taking as a standpoint the classical representation of physical reality these approaches have focused their research in trying to modify the orthodox quantum formalism in order to make sense of their (classical) metaphysical prejudices. David Bohm, reasoning that in physics we only measure `positions', concluded that QM should talk about `particles with definite trajectories'. Modifying Schr\"odinger's equation, Bohm was able to define a new (quantum) field that would guide the particles in their (space-time) trajectories ---avoiding in this way both `collapses' and the weird `interaction of probabilities'. Giancarlo Ghirardi, Alberto Rimini, and Tullio Weber also modified Schr\"odinger's (linear) equation adding a non-linear term in order to describe in a supposedly realist manner the collapse of quantum superpositions ---added, let us not forget, by anti-realists.\footnote{In turn, adding to the ramification, both interpretations were later on also discussed in terms notions, such as propensities, dispositions, properties, particles, fields, flashes, etc., which were part of other `interpretations'. All, not necessarily consistent, combinations of different interpretations have been accepted in the specialized literature as possible narratives for QM.} Very soon, even those interpretations which had begun as continuations of the general (anti-realist) positivist-Bohrian program, like for example the proposals by Hugh Everett and Bas van Fraassen, also ended up falling pray of metaphysical ``realist'' narratives. While Everett's relativist-observational proposal was reinterpreted by DeWitt ---and later on also by Oxfordians like David Deutsch, David Wallace and Simon Saunders--- in terms of a {\it multiverse} in which the branching of a multiplicity of real existing worlds would explain the process of quantum measurement as well as the efficiency of quantum computers; the Copenhagen modal interpretation proposed by van Fraassen was reinterpreted by Dennis Dieks as making reference not only to actualities (i.e., to measurement outcomes) but also to `systems' (i.e., particles) with definite valued properties. However, it is interesting to notice that even ``non-collapse'' (realist) interpretations have continued to make explicit use of the (anti-realist) measurement rule in order to address observations ---just like their collapse partners--- in completely naive empirical terms. Let us discuss these interpretations in some detail.



\subsection{The Many Worlds Narrative: The Measurement Rule as a Branching Process}

Today, the existence of many parallel worlds, similar to ours, has become a popular idea not only in films and TV series but also in science itself. Applied in String Theory, Cosmology and the Standard Model, the many worlds narrative has ---during the last decade--- also become one of the most popular between the many interpretations of QM. The so called many worlds interpretation of QM goes back to Everett's relative state interpretation\footnote{What seems very paradoxical with respect to the present Oxfordian account of Everett's ideas ---mainly due to Deutsch, Wallace and Saunders--- is the complete elimination of Everett's positivist standpoint regarding observability, prediction and anti-metaphysical commitments (section 2.1). Jefferey Babrret, who studied Everett's original texts and was responsible for the edition of his complete works \cite{Barrett12}, has repeatedly remarked that Everett's ideas have nothing to do with the present Oxfordian misuse of his name. In fact, it is very easy to see that Everett's Relative State interpretation of QM is much closer to QBism or Rovelli's Relational Interpretation than to the MWI.} which might be also considered as framing Bohr's contextual relativism in the more rigorous context of the positivist scheme (see \cite{deRondeFM18, Osnaghi09}). Taking as a standpoint Wigner's friend paradox which exposed the confronting representations of a superposition {\it before} and {\it after} an actual measurement was observed by an agent inside a lab and his friend outside, Hugh Everett attempted as a student to avoid the ``collapse'' of the quantum wave function and in this way escape the measurement problem right from the start. He was completely right to argue that the formalism never made reference to the existence of such strange physical process. Everett wanted the impossible to be done: escape the problem that positivism had introduced without abandoning the positivist understanding of physical theories ---which was actually responsible for introducing the measurement postulate in the first place. Following the path laid down by Bohr, he was ready to take a step further and explicitly replace the {\it reference} of QM from the representation of `quantum particles' to the observations made by agents of `clicks' in detectors. Everett had returned in this way to the basic empirical-positivist presupposition according to which physics does not make reference to {\it physis} (or reality) and should be regarded instead as a ``tool'' to be used by individual agents in order to predict measurement outcomes. In his own words \cite[p. 253]{BarrettByrne}: ``To me, any physical theory is a logical construct (model), consisting of symbols and rules for their manipulation, some of whose elements are associated with elements of the perceived world.'' Following the positivist program, Everett's account of QM made no {\it reference} to any picture or image beyond the observation of `clicks'. In this respect, he stressed: ``There can be no question of which theory is `true' or `real' ---the best that one can do is reject those theories which are not isomorphic to sense experience.'' Once QM was understood by Everett as making reference only to the relative observations made by agents, the `collapse' seemed to have finally disappeared. And yet, the measurement rule (or projection postulate) had not. Instead, it had changed its reference from the strange ``collapse'' of the quantum wave function to an even stranger ``branching process'' also dependent on subjects. Alike the measurement axiom introduced by Dirac and von Neumann, this ``branching'' was not meant to be interpreted in realist or metaphysical terms. As explained by Jeffrey Barrett, the hole point of Everett's (dis)solution of the measurement problem was that ``the branching'' had to be understood in purely operational terms \cite{BarrettByrne}. However, and regardless of Everett's intentions, during the early 1970s Bryce DeWitt's and Neill Graham  conceived the branching ---beyond its operational reference--- in metaphysical terms. According to DeWitt and Graham, the quantum measurement interaction described by the branching process was actually responsible for creating many different {\bf real worlds} ---mathematically represented in terms of quantum superpositions. In this way, Everett's relativist anti-metaphysical proposal was turned completely upside-down. It is then not so strange that Barrett, who had access to Everett's original notes, found written next to the passage where DeWitt presented Graham's many worlds clarification of Everett's own ideas the word ``bullshit''.\footnote{See \cite[364-6]{Barrett12} for scans of  Everett's comments.} In DeWitts' and Graham's many worlds reformulation the ``branching'' became once again ---just like the ``collapse''--- a real physical process. A process which allowed a measurement in one ---real--- world to create a myriad of parallel ---also real--- worlds. Of corse, nothing of this incredible ``creationist process'' was described by the theory.\footnote{In this respect, it is very interesting to notice that we could think of the ``branching process'' as the mirror image of the ``collapse process'' ---none of which is addressed nor explained by QM. While the collapse turns the superposition into only one if its terms, the branching goes from one single measurement into a superposition of parallel worlds.} During the 1970s, DeWitt's ideas were not taken very seriously by the physics community. A new launching was produced in the 1980s when David Deutsch applied the existence of many worlds in order to explain the efficiency of quantum computations \cite{Deutsch85}. But still, the many worlds' branching solution to the measurement problem remained as unclear as its hylomorphic collapse twin. The reason is quite simple, there is nothing in the mathematical formalism of QM which can be related to the measurement postulate on which ---of course--- both collapse and branching processes are explicitly grounded. Thus, there can be no formal representation whatsoever in the theory which explains {\it how} or {\it when} any of these amazing {\it ad hoc} physical processes really takes place. In this respect, since the theory of quanta is linear and never mentions any branching process, it becomes difficult to understand the claim by Deutsch and Wallace that the many worlds interpretation is a ``literal'' reading of the orthodox quantum formalism. Another troubling aspect of the many worlds proposal is its unclear representation of the state of affairs which seems to talk not only about particles but also about a possible future branching of worlds (see \cite{Sudbery16}). According to many worlds, QM is not describing an actual state of affairs, but instead a future state of affairs which will be only actualized {\it after} the branching has taken place. The quantum superposition seems to make {\it reference} to the many worlds created only {\it after} an actual measurement has been performed. This places the interpretation in a very difficult dilemma regarding the reference provided by the ---present and future--- representation of the theory. The many worlds interpretation seems to imply that QM makes reference only to a {\it future} state of affairs, but not to the present one. One might also wonder about the {\it reference} to `elementary particles' which, {\it before} the measurement, also seems to be described by quantum superpositions. Quantum superpositions seem to make reference before the measurement to microscopic particles, and after the measurement, to many macroscopic parallel worlds. But apart from the unclear reference of what the interpretation really talks about, there are many questions related to the branching process which do not seem to find a convincing answer. What about the {\it interference} of probabilities and superpositions? If quantum probability is understood as (epistemic) `degrees of belief', how can they (really) interact? And what about entanglement? Does it amount to an interaction of one superposition of many worlds with another superposition of many worlds... or is it elementary particles? Where does this interaction takes place? And when? And what would it mean that a world interferes or interacts with another world? How does one actual world affect another parallel one? How can the branching and interference of worlds be tested experimentally? And by the way, how many worlds exactly are created in each branching? How many of these worlds really exist in our multiverse?\footnote{In recent interviews the many worlds followers have been confronted to some of these questions. Wallace's \cite{WallaceCT} answer is that: ``It is hard to define exactly because this branching process is not precise, but to put a number out of the air $10^{10^{100}}$, so 10 to the number of particles in the Universe.'' We might point out that the acknowledgment that ``the branching process is not precise'' in an interpretation which attempts to describe ``literally'' the quantum formalism seems, to say the least, very unsatisfactory. Another question which immediately pops up is how did Wallace compute the number,  $10^{10^{100}}$?} How is all this actually represented by the mathematical formalism of QM?

\subsection{Dieks' Modal Narrative: The Measurement Rule as a Semantic Rule} 

Another realist ``non-collapse'' interpretation is Dennis Dieks' modal version of Bas van Fraassen's (anti-realist) modal interpretation. Contrary to van Fraassen, Dieks interpretation was conceived in order to make sense of QM in terms of `systems' with definite valued `properties' ---going in this way beyond observed actualities. According to him, the attempt was to do so staying close to the orthodox quantum formalism without adding anything ``by hand'' (for a detailed analysis see \cite{deRonde11, Vermaas99}). Something that, according to Dieks, implied leaving behind the existence of ``collapses''.
\begin{quotation}
\noindent {\small ``Collapses constitute [...] a process of evolution that conflicts with the evolution governed by the Schr\"{o}dinger equation. And this raises the question of exactly when during the measurement process such a collapse could take place or, in other words, of when the Schr\"{o}dinger equation is suspended. This question has become very urgent in the last couple of decades, during which sophisticated experiments have clearly demonstrated that in interaction processes on the sub-microscopic, microscopic and mesoscopic scales collapses are never encountered.'' \cite[p. 120]{Dieks10}}
\end{quotation}

Dieks modal narrative follows van Fraassen's semantic account of theories \cite{Dieks91} where ``an uninterpreted theory is identified with the class of its models, in the sense of abstract model theory. [...] To make an empirical theory of it, we have to indicate how empirical data can be embedded in the models.'' In particular, in order ``[t]o obtain quantum mechanics as an empirical theory, we have to specify the links with observation, by means of `interpretation rules'.'' It is at this point that Dieks rephrases the measurement rule in his own terms:
\begin{quotation}
\noindent {\small ``Consider a state vector representing a composite system, consisting of
an object system and the remainder of the total system. The total state
vector will almost always have one unique bi-orthonormal decomposition:
\begin{center}
$| \Psi \rangle = \sum c_k  |\psi_k \rangle |R_k \rangle $ \ \ \ \ \ \ \ \ \ \ \ \  (2)
\end{center}
where the $|\psi_k \rangle$ refer to the object system and the $|R_k \rangle$ to the rest of the
system; $\langle \psi_i | \psi_j \rangle$ and $\langle R_i | R_j \rangle = \delta_{ij}$. I now propose the following {\it semantical
rule:} {\bf As soon as there is a unique decomposition of the form (2), the
partial system represented by the $|\psi_k \rangle$, taken by itself, can be described as
possessing one of the values of the physical quantity corresponding to the set $\{ |\psi_k \rangle \}$. The probabilities for the various possibilities to be realized are given by $| c_k|^2$.}'' \cite[p. 1406]{Dieks89}}
\end{quotation}
The attentive reader might have already realized that the {\it semantic rule} ---also called by Dieks  {\it interpretational rule}--- is nothing essentially different from the measurement rule or projection postulate. Independently of the introduction of the Schmidt decomposition, Dieks' rule ends up doing exactly the same job as its predecessors, namely, to magically ``bridge the gap'' between quantum superpositions and single measurement outcomes. The tension present in Dieks' empiricist-realist program becomes then evident. At the end of the day, his narrative ---in a completely Bohrian fashion--- retains two contradictory claims. On the one hand,  that QM should be understood as representing physical reality in terms of elementary quantum particles (or systems) with definite valued properties, and on the other, that QM should be used ---through the application of his interpretational rule--- as a ``tool'' in order to predict observations (or measurement outcomes). Even though Dieks begins by arguing that the mathematical formalism of the theory should be interpreted ``in realist terms'' as making reference to the microscopic world and {\it ad hoc} rules should be rejected, he immediately reintroduces the measurement postulate now renamed as the a ``semantic'' or ``interpretational'' rule. Even though Dieks denies the existence of ``collapses'', he anyhow applies his semantic rule in order to justify the appearance of single `clicks' in detectors. In this way, the mathematical formalism is dissected in two level, a supposedly realist level which is unclearly related to `systems' with definite valued `properties' (i.e., to quantum particles), and an anti-realist level which through the {\it ad hoc} introduction of the interpretational rule does the dirty job of providing an instrumentalist account of measurement outcomes.  

It is important to understand that the reduced states arising from the Schmidt decomposition ---to which Dieks applies his interpretational (measurement) rule--- are in fact {\it improper mixtures} which, just like in the case of quantum superpositions, cannot be interpreted in terms of ignorance \cite[Chap. 6]{DEspagnat76}.\footnote{This point is recognized explicitly by Dieks in many occasions. See, for example: \cite[p. 1407]{Dieks89}.} Thus, absolutely nothing has been gained by shifting the reference of the orthodox quantum formalism from one single system to two correlated ones. Notwithstanding the fact that one might wonder about the meaning of a theory that would describe `composite systems' but fails to describe `single systems', the attempt to provide a consistent interpretation without the addition of {\it ad hoc} rules is given up right from the start. Following the same strategy, Dieks and Pieter Vermaas \cite{VermaasDieks95} have argued that one needs to name the {\it reduced state} in two different ways. First, as a ``mathematical state'' which cannot be interpreted in terms of ignorance ---because of the mathematical formalism---, and then as a ``physical state'' which can be interpreted in terms of ignorance ---simply because it is now called ``physical''. The solution is to shift the naming of the same state depending on the context of application.\footnote{In fact, this linguistic duality had been already proposed by van Fraassen who distinguishes between {\it dynamical states} and {\it value states}. What van Fraassen does not do, is to provide a realist interpretation of such a distinct naming.} The fact that the properties of systems do not behave according to the classical intuition \cite{Bacciagaluppi95, Clifton95b, Vermaas97, RFD14} is resolved by Dieks in a completely Bohrian fashion: the modal interpretation does not provide the expected representation of systems simply because QM is not classical.\footnote{This argument was first used by Bohr who argued in 1927 that ``quantum jumps'' could not be represented. As recalled by Heisenberg \cite[p. 74]{Heis71}, Bohr's reply to Schr\"odinger's criticisms related exclusively to the limits imposed by QM:  ``What you say is absolutely correct. But it does not prove that there are no quantum jumps. It only proves that we cannot imagine them, that the representational concepts with which we describe events in daily life and experiments in classical physics are inadequate when it comes to describing quantum jumps. Nor should we be surprised to find it so, seeing that the processes involved are not the objects of direct experience.'' As Bohr \cite[p. 701]{Bohr35} would argue: ``The impossibility of a closer analysis of the reactions between the particle and the measuring instrument is indeed no peculiarity of the experimental procedure described, but is rather an essential property of any arrangement suited to the study of the phenomena of the type concerned, where we have to do with a feature of [quantum] individuality completely foreign to classical physics.''}  
\begin{quotation}
\noindent {\small ``It is obvious then that our proposed semantical rule, together with the formalism of quantum mechanics, can lead to consequences which are very much at variance with classical intuition. In classical physics the properties of all partial systems together always completely determine the properties of a total system. By contrast, quantum theory ---with the interpretation as discussed above--- sometimes associates with a composite system a description that contains more than just the properties of the partial systems by themselves.'' \cite[p. 1408]{Dieks89}}
\end{quotation}

To sum up. Just like in the Bohrian scheme analyzed in detail in \cite{deRonde20b}, there is in Dieks' interpretation a dualistic (inconsistent) account of QM. On the one hand, there is an instrumentalist (anti-realist) account of the theory provided through rules which only make reference to measurement outcomes; and on the other, a metaphysical (realist) account which makes reference to systems with properties. But while at the metaphysical level ``collapses'' are rejected by claiming that one should not accept {\it ad hoc} additions, at the empirical level the measurement postulate is accepted as part of the theory. But even worse, not only the orthodox mathematical formalism fails to provide a consistent account of `quantum systems' as possessing definite valued properties, the interpretation fails to produce any intuitive ({\it anschaulicht}) insight. The addition of quantum particles to the interpretation seems to play no other role than supporting a fictional narrative with no clear explanatory power nor contact with the mathematical formalism of the theory.



\section{The Missing Link Between Mathematics and Physical Concepts}

The positivist scheme fails to provide any scientific methodology in order to address the link between the mathematical formalism and the conceptual framework of a given theory. Not only observations ---which are considered as primary {\it givens} of experience--- are independent of conceptual or categorical constraints, also the mathematical formalism ---which is developed in order to account for observations--- is completely independent of metaphysical concepts. According to orthodoxy, physicists must create mathematical models in order to account for what experimentalists observe in the lab. That's it! Conceptual (or metaphysical) representation plays no role whatsoever within the construction of an empirical adequate theory. This becomes explicitly recognized by the role assigned to `interpretations' as `fictional narratives' or `stories' which are introduced (by philosophers) {\it a posteriori} from the construction of theories (by physicists). In tune with the postmodern {\it Zeitgeist} of the 20th Century, interpretations have no other fundament than the wishful hopes and beliefs of a metaphysically inclined community which still attempts to talk about a representation of reality beyond what we actually observe here and now. As remarked by David Deutsch: 
\begin{quotation}
\noindent {\small ``[Postmodernism itself] is a narrative that resists rational criticism or improvement, precisely because it rejects all criticism as mere narrative. Creating a successful
postmodernist theory is indeed purely a matter of meeting the criteria
of the postmodernist community ---which have evolved to be complex,
exclusive and authority-based. Nothing like that is true of rational ways
of thinking: creating a good explanation is hard not because of what
anyone has decided, but because there is an objective reality that does
not meet anyone's prior expectations, including those of authorities.
The creators of bad explanations such as myths are indeed just making
things up. But the method of seeking good explanations creates an
engagement with reality, not only in science, but in good philosophy
too ---which is why it works, and why it is the antithesis of concocting
stories to meet made-up criteria. Although there have been signs of improvement since the late twentieth
century, one legacy of empiricism that continues to cause confusion,
and has opened the door to a great deal of bad philosophy, is the idea
that it is possible to split a scientific theory into its predictive rules of
thumb on the one hand and its assertions about reality (sometimes
known as its `interpretation') on the other. This does not make sense,
because ---as with conjuring tricks--- without an explanation it is
impossible to recognize the circumstances under which a rule of thumb
is supposed to apply. And it especially does not make sense in fundamental
physics, because the predicted outcome of an observation is
itself an unobserved physical process.'' \cite[p. 1408]{Deutsch04}}
\end{quotation}
The problem present within the positivist scheme is that there is no methodology allowing to discuss in scientific terms ---beyond wishful thinking and subjective preferences--- if an interpretation provides (or not) a good explanation of the mathematical formalism (see \cite{Chakravartty17, French11}). It is the complete lack of any objective criteria to distinguish between `good' and `bad' interpretations which, in turn, has allowed the wild introduction of fictional notions, principles, rules and pictures with no contact whatsoever with the mathematical formalism or experience. Let's give a few examples. There is no mathematical element of the formalism describing a `particle' in Bohmian mechanics. The same occurs in the many worlds interpretation where there is no description of what a `world' {\it is} or how the branching process could be described or computed according to the theory. Just like in the case of `collapses' and `quantum particles' these notions help creating a story which does not relate to the mathematical formalism they ---supposedly--- attempt to represent. The experimental testing of the added concepts, an essential requirement for any consistent physical notion, is also missing. In fact, even those interpretations which attempt to respect the mathematical constraints to physical concepts have not taken very seriously their own research, and instead of critically addressing their findings they have continued to add new principles and notions which rather than clarify our understanding simply add to the general confusion. We find two paradigmatic examples in the just discussed ``non-collapse'' interpretations. While many worlds interpretations, when confronted to the problem of interpreting the Born rule, have ended up following the (anti-realist) Bayesian interpretation of probability; Dieks modal interpretation, when confronted to different no-go theorems which explicitly show that `quantum systems' cannot be regarded as possessing definite valued properties, has turned into an extreme form of (anti-realist) relativist scheme, namely, perspectivalism \cite{BeneDieks02, Dieks19}. Both narratives have also made explicit use of the principle of decoherence in order to address the basis problem. But this principle, as it has been already acknowledged in the philosophical literature, fails to solve any of the problems it was designed to. Subverting explaination, decoherence has created a new form of justification known in the literature as ``FAPP'', something which amounts to Professors yelling at students ``shut-up and calculate!''  

Regardless of the ---apparent--- generalized turmoil and disputes between the many interpretations of QM, there is a common methodology which has allowed to construct these stories. The Bohrian methodology ---as we might call it--- works as follows. First dissect the theory in two parts. One which provides the operational empirical content, and another part which is related to a ---supposedly realist--- metaphysical narrative. While the the first might be read as an instrumentalist algorithm which allows to compute an ``observed'' or ``internal reality'', the latter can be related to a representation of an ``external reality''. At this point you are free to add the notions of your preference: systems, worlds, particles, branchings, collapses, histories, flashes, propensities or whatever you come up with. It doesn't really matter for the representation will remain essentially unrelated to the mathematical formalism or ---even--- experience. Even if the metaphysical story becomes unclear there are many tricks you can use to escape the need of explanation: you can blame it to theory of quanta itself by shouting as loud as you can ``It is quantum!'', or you can also add {\it ad hoc} principles which do not necessarily explain anything but will either naturalize the problem ---like, for example, the principles of correspondence and complementarity---  or shift the attention to another problem ---like in the case of decoherence. This methodology is explicit in Dieks' {\it interpretational rule} which has a purely operational reference but no link whatsoever to the way in which `quantum systems' and `properties' behave. The same happens with the role played by quantum superpositions in the many worlds interpretation. While projection operators are interpreted as related to (real) `properties' of elementary particles present in different worlds, the numbers which accompany the states are interpreted as (unreal) `degrees of beliefs' of agents (or subjects). To sum up. Dissect the theory in two, one part in order to make predictions and another part to talk about reality. Add as many unexplained notions as you can. If there is something that does not fit your story simply add something else which is less understood. In order to make the narrative completely invulnerable to scientific criticism it is of outmost importance that ---just like in the case of quantum jumps, collapses, branching, complementarity, correspondence, etc.--- you do not relate the added ``explanation'' to an experimental procedure. There should exist no test allowing to conclude if what the interpretation says is tenable (or not) in a given situation. If you do this with ``studied ambiguity'' ---as Deutsch has characterized it \cite{Deutsch04}--- the confusion will grow in such a way that the reader will not be able to follow you and at some point he will start doubting herself about her own capability of understanding. If nothing of this works, just blame the theory itself for its weirdness and failures. Just repeat the mantra: ``It is quantum!''

\smallskip 

In contraposition to the positivist anti-metaphysical scheme, we have argued repeatedly that metaphysics does play an essential role within scientific theories. Metaphysics is not a story about the unobservable realm. Metaphysics involves the creation of nets of interrelated concepts which allows us to {\it represent} reality and experience in a systematic manner. Observability in physics has nothing to do with ``common sense'', it is instead a {\it praxis} derived from the formal-conceptual unity provided by a theory.\footnote{For a critical analysis of the role of observation in physical theories see \cite{Deutsch04, Maudlin19}.} As Tian Yu Cao makes the point: 
\begin{quotation}
\noindent {\small``The old-fashioned (positivist or constructive empiricist) tradition to the distinction between observable and unobservable entities is obsolete. In the context of modern physics, the distinction that really matters is whether or not an entity is cognitively accessible by means of experimental equipment as well as conceptual, theoretical and mathematical apparatus. If a microscopic entity, such as a W-boson, is cognitively accessible, then it is not that different from a table or a chair. It is clear that the old constructive empiricist distinction between observables and nonobservables is simply impotent in addressing contemporary scientific endeavor, and thus carries no weight at all. If, however, some metaphysical category of microscopic entities is cognitively inaccessible in modern physics, then, no matter how basic it was in traditional metaphysics, it is irrelevant for modern metaphysics.''  \cite[pp. 64-65]{Cao03}}
\end{quotation}
It is only through the interrelation of formal-conceptual representations that theoretical experience becomes possible. As Einstein remarked to a young Heisenberg: ``It is only the theory which decides what can be observed.'' Some decades later Heisenberg \cite[p. 264]{Heis73} himself would remark that: ``The history of physics is not only a sequence of experimental discoveries and observations, followed by their mathematical description; it is also a history of concepts. For an understanding of the phenomena the first condition is the introduction of adequate concepts. Only with the help of correct concepts can we really know what has been observed.'' The creation of concepts and representations is thus essential to the very possibility of considering scientific experience. As Cao \cite[p. 65]{Cao03} continues to explain: ``An important point here is that metaphysics, as reflections on physics rather than as a prescription for physics, cannot be detached from physics. It can help us to make physics intelligible by providing well-entrenched categories distilled from everyday life. But with the advancement of physics, it has to move forward and revise itself for new situations: old categories have to be discarded or transformed, new categories have to be introduced to accommodate new facts and new situations.'' This is an essential point of disagreement with Bohr's doctrine of classical concepts. On this point, Einstein had already warned us about the dangers of dogmatism: 
\begin{quotation}
\noindent {\small ``Concepts that have proven useful in ordering things easily achieve such an authority over us that we forget their earthly origins and accept them as unalterable givens. Thus they come to be stamped as `necessities of thought,' `a priori givens,' etc. The path of scientific advance is often made impossible for a long time  through such errors. For that reason, it is by no means an idle game if we become practiced in analyzing the long common place concepts and exhibiting those circumstances upon which their justification and usefulness depend, how they have grown up, individually, out of the givens of experience. By this means, their all-too-great authority will be broken. They will be removed if they cannot be properly legitimated, corrected if their correlation with given things be far too superfluous, replaced by others if a new system can be established that we prefer for whatever reason.'' \cite{Howard10}}
\end{quotation}
It is from this realist understanding of physical theories, as providing many different representations of a state of affairs, that we have argued in \cite{deRonde18} that the (anti-realist) measurement problem should be replaced by what we have termed the superposition problem.  

\smallskip 
\smallskip 

\noindent {\it {\bf Superposition Problem:} Given a situation in which there is a quantum superposition of more than one term, $\sum c_i \ | \alpha_i \rangle$, and given the fact that each one of the terms relates through the Born rule to a meaningful operational statement, the question is how do we conceptually represent this mathematical expression? What are the physical concepts that relate to each one of the terms in a quantum superposition?}

\smallskip 
\smallskip 

\noindent Just in the same way that the principles of existence, non-contradiction and identity have played a double role defining on the one hand the notion of `entity' in the actual mode existence, and on the other, classical logic itself; quantum superpositions must also need to find a conceptual framework which unlocks its physical meaning. Heisenberg's original mathematical formulation of QM was developed from empirical findings which escaped the classical representation, it is from the mathematical formalism and its operational content that we should derive a consistent and coherent conceptual scheme. It is in this context that the notion of `elementary particle' together with the notion of `event' have played the roles of metaphysical and empirical obstructions restricting the possibilities of a conceptual development; according to Wolfgang Pauli \cite[p. 193]{Laurikainen98}, ``the most important and extremely difficult task of our time to work on the elaboration of a new idea of reality.'' In this respect, Pauli also remarked that \cite[p. 126]{Pauli94}: ``We [should] agree with P. Bernays in no longer regarding the special ideas, which Kant calls synthetic judgements {\it a priori}, generally as the pre-conditions of human understanding, but merely as the special pre-conditions of the exact science (and mathematics) of his age.'' What we desperately need for QM is a new non-classical conceptual scheme which is consistently linked to the mathematical formalism and allow us to think ---without {\it ad hoc} rules and principles--- in a truly consistent manner about the states of affairs and experience described by the theory. In this respect, it is of outmost importance to go back to the start of the quantum voyage and recall the way in which the mathematical formalism of QM was actually developed from a specific field of intensive experience.

\section{What Do We Actually Observed in the Lab: A `Click' or a `Pattern'?} 

The idea that QM should be able to predict single `clicks' in detectors is a naive empiricist prejudice that was explicitly introduced to the theory by Paul Dirac. Leaving behind the subversive spirit of the positivist principle which had allowed Heisenberg to escape classical atomism and develop matrix mechanics,\footnote{As Heisenberg begun his foundational paper of 1925: ``The present paper seeks to establish a basis for theoretical quantum mechanics founded exclusively upon relationships between quantities which in principle are observable. It was this same principle which was also kernel for Einstein's development of special relativity and his criticism of the notion of simultaneity.} Dirac restricted the definition of observability to a binary understanding of {\it certainty}.\footnote{As explained by Asher Peres See \cite[p. 66]{Peres93}: ``The simplest observables are those for which all the coefficients $a_r$ are either 0 or 1. These observables correspond to tests which ask {\it yes-no questions} (yes = 1, no = 0).''} 
\begin{quotation}
\noindent {\small ``We now make some assumptions for the physical interpretation of the theory. {\it If the dynamical system is in an eigenstate of a real dynamical variable $\xi$, belonging to the eigenvalue $\xi'$, then a measurement off $\xi$ will certainly give as result the number $\xi'$. Conversely, if the system is in a state such that a measurement of a real dynamical variable $\xi$ is certain to give one particular result (instead of giving one or other of several possible results according to a probability law, as is in general the case), then the state is an eigenstate of $\xi$ and the result of the measurement is the eigenvalue of $\xi$ to which this eigenstate belongs.}'' \cite[p. 35]{Dirac74} (emphasis in the original)}
\end{quotation} 
This distinction introduced within the theory ---between {\it certain} and {\it uncertain} observations--- was not to be found in the the field of phenomena from which the original mathematical formalism had been actually developed. In fact, since its origin, the theory of quanta was related to the line intensity patterns of radiation and frequency problems where single measurement outcomes where simply not considered.\footnote{This is analogous to the way in which the measurement of temperature does not address the velocity of single particles.}

The construction of the theory of quanta took place during the first quarter of the 20th Century. As described by Heisenberg \cite[p. 3]{Heis58}: ``The origin of quantum theory is connected with a well-known phenomenon, which did not belong to the central parts of atomic physics. Any piece of matter when it is heated starts to glow, gets red hot and white hot at higher temperatures. The color does not depend much on the surface of the material, and for a black body it depends solely on the temperature. Therefore, the radiation emitted by such a black body at high temperatures is a suitable object for physical research; it is a simple phenomenon that should find a simple explanation in terms of the known laws for radiation and heat.'' Regardless of the simplicity of the problem, the difficulties were immense. It had become impossible to find ---by following classical presuppositions--- a suitable theoretical model that would account for the experience observed in the lab. This was until in the year 1900, Max Planck was finally able to operationally solve what was known as the ``ultraviolet catastrophe'' by introducing the famous {\it quantum of action} in the Rayleigh-Jeans law of intensive radiation for black bodies. It took 25 years for physicists to reach a closed mathematical formalism that would allow the theory of quanta to account in a consistent manner for what was actually observed in the lab. Heisenberg was able to develop matrix mechanics following two ideas, first, to leave behind the classical (metaphysical) notion of particle-trajectory, and second, to take seriously Ernst Mach's positivist rule according to which a theory should only make reference to what is actually observed in the lab. So what was observed? The answer is well known to any experimentalist: a spectrum of line intensities. This is what was described by the tables of data that Heisenberg  had attempted to mathematically model and finally led him ---with the help of Max Born and Pascual Jordan--- to the development of the first mathematical formulation of the theory of quanta, namely, Quantum Mechanics. As he would himself recall:  
\begin{quotation}
\noindent {\small ``In the summer term of 1925, when I resumed my research work at the University of G\"ottingen ---since July 1924 I had been {\it Privatdozent} at that university--- I made a first attempt to guess what formulae would enable one to express the line intensities of the hydrogen spectrum, using more or less the same methods that had proved so fruitful in my work with Kramers in Copenhagen. This attempt lead me to a dead end ---I found myself in an impenetrable morass of complicated mathematical equations, with no way out. But the work helped to convince me of one thing: that one ought to ignore the problem of electron orbits inside the atom, and treat the frequencies and amplitudes associated with the line intensities as perfectly good substitutes. In any case, these magnitudes could be observed directly, and as my friend Otto had pointed out when expounding on Einstein's theory during our bicycle tour round Lake Walchensee, physicists must consider none but observable magnitudes when trying to solve the atomic puzzle.'' \cite[p. 60]{Heis71}}
\end{quotation}
Of course, operational observations and their algorithmic prediction were not enough in order to constitute a unified, consistent and coherent account of phenomena. As Heisenberg \cite[p. 264]{Heis73} would make the point: ``For an understanding of the phenomena the first condition is the introduction of adequate concepts. Only with the help of correct concepts can we really know what has been observed.'' And this is the essential gap between the operational prediction of observations in the lab and the conceptual understanding of physical phenomena. 
\begin{quotation} 
\noindent {\small  ```Understanding' probably means nothing more than having whatever ideas and concepts are needed to recognize that a great many different phenomena are part of coherent whole. Our mind becomes less puzzled once we have recognized that a special, apparently confused situation is merely a special case of something wider, that as a result it can be formulated much more simply. The reduction of a colorful variety of phenomena to a general and simple principle, or, as the Greeks would have put it, the reduction of the many to the one, is precisely what we mean by `understanding'. The ability to predict is often the consequence of understanding, of having the right concepts, but is not identical with `understanding'.'' \cite[p. 63]{Heis71}} \end{quotation}
Following the representational realist program, a theory must be considered as a unity of meaning and sense, a wholeness constructed in both mathematical and metaphysical (or conceptual) terms which allows us to {\it represent} a specific field of phenomena. It is exactly this consistent and coherent wholeness which provides the missing structural link between a mathematical formalism, a net of concepts and a field of phenomena. This unity, in order to be consistent, must also allow us to think beyond particular mathematical reference frames or the specific viewpoint of empirical agents. This is a necessary condition for any consistent and coherent representation of physical reality which Einstein stressed repeatdly. This requirement also provides a structural link between mathematics and concepts. In particular, it is exposed in the intrinsic relation between operational invariance and conceptual objectivity to which we now turn our attention.

\section{From Operational Invariance to Conceptual Objectivity} 

Taking as a standpoint that in physical theories concepts are essential for observation, that they are relational constructs which depend on their mutual interrelation as a whole and that this whole must be consistently related to mathematical formalisms, it is possible to provide a completely different approach to the orthodox empirical-positivist understanding of theories. This approach, which we have termed {\it representational realism} \cite{deRonde16b}, goes back not only to the writings of Einstein, Pauli and Heisenberg but also to the original ancient Greek meaning of physics. According to this understanding, physical theories provide a unified, consistent and coherent formal-conceptual invariant-objective representation of states of affairs and experience. That is what theories do. In the case of QM, what we are missing is the conceptual framework in order to understand what we observe. As we have argued elsewhere \cite{deRondeMassri17}, the key to develop such an objective conceptual representation can be found in the operational-invariant structure of the theory itself. As Max Born \cite{Born53} reflected: ``the idea of invariant is the clue to a rational concept of reality, not only in physics but in every aspect of the world.'' Indeed, in physics {\it invariants} ---quantities having the same value for any reference frame--- allows us to determine what can be considered ---according to a mathematical formalism--- {\it the same} independently of the particular choice of a reference frame. The transformations that allow us to consider the physical magnitudes from different frames of reference have the property of forming a group. While in the case of classical mechanics invariance is provided via the Galilei transformations and in relativity theory via the Lorentz transformations, in QM the invariant content of the theory is brought by no other than Born's famous rule.

\smallskip
\smallskip

\noindent {\it
{\bf Born Rule:} Given a vector $\Psi$ in a Hilbert space, the following rule allows us to predict the average value of (any) observable $P$. 
$$\langle \Psi| P | \Psi \rangle = \langle P \rangle$$
This prediction is independent of the choice of any particular basis.}

\smallskip
\smallskip

\noindent This rule, which gives the operational-invariant content to the theory, provides the guiding line to develop an objective representation without the need to refer to objects. An objectivity without objects.\footnote{As proposed in \cite{deRonde16} there is a natural extension of what can be considered to a {\bf Generalized Element of Physical Reality:} {\it If we can predict in any way (i.e., both probabilistically or with certainty) the value of a physical quantity, then there exists an element of reality corresponding to that quantity.}} While invariance allows to detach the mathematical representation from particular mathematical frames, it is {\it objectivity} ---understood in abstract terms--- which allows us to detach the conceptual representation from the particular observations made by (empirical) subjects. The detachedness of the subject is essential for physics to refer to reality. As Einstein remarked, observation cannot change the representation of the state of affairs which is always {\it prior} to the very possibility of experience. The moon must have a position independently of us observing it or not. The notions of invariance and objectivity are intrinsically related, the first being the mathematical counterpart of the second, and the latter being the conceptual counterpart of the first (see for a detailed analysis \cite{deRondeMassri17}). It is this interrelation between mathematics and concepts which allow us to create a formal-conceptual representational {\it moment of unity} which is able to subsume a multiplicity of phenomena independently of reference frames or --even--- particular observations. Following this line of reasoning, it becomes natural to understand Born's rule as providing objective {\it intensive information} of a (quantum) state of affairs instead of subjective information about measurement outcomes (see for a detailed analysis and discussion \cite{deRonde16, deRondeFreytesSergioli19}). It is in this way that the pieces of the puzzle begin to fall into place. Intensity becomes not only the phenomenological standpoint of development of the mathematical formalism of the theory but also the objective support of a new conceptual scheme which must go beyond the classical binary representation of classical physics.  

In several papers \cite{deRondeMassri18a, deRondeMassri18b}, we have presented a formal-conceptual representation which provides explicit relationships between Heisenberg's mathematical formalism ---without measurement rules--- and a specially suited set of (non-classical) concepts. While projection operators are related to the notion of {\it intensive power}, their potentia is computed via the Born rule in an invariant manner. This non-classical representation of reality implies a radical shift from a {\it binary} picture of `systems' composed by definite valued `properties', to an {\it intensive} account of `powers' with definite `potentia'. The Born rule is then understood as quantifying the mode of existence of powers (i.e., the potentia). Thus, unlike properties conceived in binary terms (as strictly related to the values 0 or 1), quantum powers extend the realm of existence to an intensive level in which any potentia pertaining to the interval [0,1] is accepted as a {\it certain intensive value}. In this way ---escaping from empirical and metaphysical prejudices--- certainty is not restricted to the actual realm. The potentia $p = 0.764$ is not the measure of ignorance of an agent but the computation of an intensive value which can be tested operationally. Of course, this move has deep consequences not only for the understanding of the formalism but also for the experience discussed by the theory. The intensive valuation of powers has allowed us not only to escape the reference to measurement outcomes ---and consequently, to avoid the measurement problem--- but also to bypass Kochen-Specker contextuality. In this latter respect, we have derived a non-contextual intensive theorem  \cite[Theo. 4]{deRondeMassri18a} which, through the provision of a global (intensive) valuation of all projection operators, allows to understand contexts (or bases) in objective terms \cite{deRondeMassri17}. 

From a metaphysical viewpoint, the existence of quantum powers implies a potential realm of existence. Consequently we obtain a shift from a (classical-binary) representation in terms of an {\it Actual State of Affairs} (ASA) to a (quantum-intensive) representation in terms of a {\it Potential State of Affairs} (PSA).\footnote{It is important to stress that the potential mode of existence to which we refer is completely independent of the actual realm and should not be understood in teleological terms as referring to the future actualization of measurement outcomes \cite{deRonde17}.} As explained in detail in \cite{deRondeMassri18a}, intuitively, we can picture a PSA as a table,
\[
\Psi:\mathcal{G}(\mathcal{H})\rightarrow[0,1],\quad
\Psi:
\left\{
\begin{array}{rcl}
P_1 &\rightarrow &p_1\\
P_2 &\rightarrow &p_2\\
P_3 &\rightarrow &p_3\\
  &\vdots&
\end{array}
\right.
\]

\noindent Our objective representation in terms of a PSA constituted by powers with definite potentia (computed via the Born rule) allows us to understand Kochen-Specker contextuality as an epistemic feature of the theory which provides a constrain to the simultaneous measurement of incompatible powers ---avoiding relativist choices which explicitly change the representation of (quantum) reality. This aspect of contextuality is also common to classical physics. Even though some powers are {\it epistemically incompatible} (i.e., they require mutually incompatible measurement set ups in order to be observed) they are never {\it conceptually incompatible} since they can be all defined to exist simultaneously through the {\it Global Intensive Valuation} determined via the Born rule. The PSA, constituted by the set of potentially existent quantum powers, is in this respect completely objective ---i.e. detached from any particular basis or observation made by an agent. Notice that this is completely analogous to the way in which the notion of ASA, as constituted by sets of actually definite valued properties, is also regarded as objective. Objective means in this case that there exists a coherent global representation of a state of affairs which is consistent with the multiple observations of phenomena and independent of the choice of the context (or basis). The main difference between an ASA and a PSA regards their different {\it conditions of objectivity}. While in the classical case an ASA is defined in terms of a set of systems with definite valued properties; in the quantum case the PSA is defined in terms of powers with definite potentia. Our approach shows explicitly a path to derive a conceptual framework which matches consistently the mathematical formalism of the theory providing at the same time an intuitive grasp of the experience involved by it. In turn, this new (non-classical) conceptual framework has allowed us to reconsider some essential notions like quantum superpositions and entanglement from a completely new objective viewpoint \cite{deRondeMassri18b, deRondeMassri19b}.

\section{Measuring Quantum Superpositions}

Given a representation of a state of affairs as described by a theory we can imagine the possible thought-experiences implied by it. This specific type of physical thinking is of course not restricted by what we have observed nor by our technical capabilities, it is only constrained by the mathematical and conceptual schemes of representation. A physicist can derive conclusions from a theory without ever observing or measuring anything. That is the true power of physics. One which allows us to advance into future developments through theoretical experience as provided by {\it Gedankenexperiments}. Einstein's and Sch\"odinger's famous thought-experiences from 1935 are good examples of the way in which physicists can go far beyond the technical restrictions of their epoch and conclude the existence of fantastic ---never before measured or observed--- phenomena like quantum entanglement. There are countless examples in the history of physics which show this same amazing theoretical capacity of predicting things that were never before observed nor even imagined. Of course, this does not mean that experimental testing should be left aside ---as some contemporary influential physicists are arguing today.\footnote{Gerard t' Hooft \cite{Hooft01} has argued that: ``Working with long chains of arguments linking theories to experiment, we must be able to rely on logical precision when and where experimental checks cannot be provided.'' Following the same line of reasoning Steven Weinberg has gone even further claiming that: ``I think 100 years from now this particular period will be remembered as a heroic age when theorists cut themselves temporarily free from their experimental underpinnings and tried and succeeded through pure theoretical reasoning to develop a unified theory of all the phenomena of nature.'' More recently, Richard Dawid has also argued in favor of considering non-empirical arguments in order to justify mathematical theories \cite{Dawid13}.} Experimental evidence is essential to physics in order to secure a consistent link between any theory and what actually happens in the here and now. Without empirical testing physical theories loose their compass, following mirages, lost and disoriented, they become an easy pray for instrumentalism and dogmatic metaphysics. As Einstein \cite[p. 175]{Dieks88a} would also make the point, even though measurement and observability cannot play an internal role within a theory, ``the only decisive factor for the question whether or not to accept a particular physical theory is its empirical success.'' It is only measurement which allows us to connect the detached experience coming from a theory with the {\it hic et nunc} observation of a phenomena in a lab, providing in this way the necessary link between (objective) theoretical representation and (subjective) empirical observability. 

While the field of thought-experience is strictly limited by the theory, observation is a purely subjective conscious action which cannot be theoretically represented. It is `measurement' which stands just in the middle between the objective representation of physical experience and the subjective {\it hic et nunc} empirical observation itself. Measurements, at least for the realist, are conscious actions performed by human subjects which are able to select, reproduce  and understand a specific type of phenomenon. This is of course a very complicated process created by humans which interrelates practical, technical and theoretical knowledge. Anyone attempting to perform a measurement must be able to think about a specific problem, she must be also able to construct a measurement arrangement, she must be capable to analyze what might be going on within the process, and finally, she must be qualified to observe, interpret and understand the phenomenon that actually takes place, {\it hic et nunc}, when the measurement is actually performed. All these requirements imply human capacities and, in particular, consciousness. The table supporting the measurement set-up does not understand what complicated process is taking place above itself. Tables and chairs cannot construct a measuring set-up. The chair that stands just beside the table cannot observe a measurement result, and the light entering the lab through the window cannot interpret what is going on. It is only a conscious (empirical) subject (or agent) who is capable of performing a measurement. As any physicist who has ever entered a lab knows very well, these actions have nothing to do with ``common sense''. 

Our theoretical standpoint allows us to reject right from the start not only the (dogmatic metaphysical) existence of quantum ``collapses'' and ``jumps'' but also the (naive empirical) requirement ---imposed by the 20th Century positivist understanding of theories--- to introduce the measurement rule (or projection postulate) within the theory. It is only adequate concepts which can allow us to understand an observation for it is only the theory which decides what can be observed. That is the reason why, when learning physics as a graduate student, you first learn the theory, and only then you go to the lab ---not the other way around. It makes no sense to try to measure an `electromagnetic field' if you have not yet learned what is the mathematical and conceptual representation of a `field'. But even having learned the theoretical definition of a `field', theories do not come with a user's manual telling us how to measure them. Every physicist has experienced  as a student the abysm of entering a lab after finishing a course in classical mechanics or  electromagnetism. As a student who has just learned a theory you simply have no clue what to do in a lab in order to test the theory you just learned in the classroom. It requires a lot of technical skills and knowledge ---that you need to learn in a lab--- in order to be capable to test something the theory implies. Of course, the theory must provide the conditions for restricting the possibilities of what can be measured. And in this respect, the most important qualitative link is provided by physical concepts. A physical concept must be able to account for the operational conditions under which it becomes possible to test experimentally its own consistency. As famously remarked by Einstein when discussing about the definition of the concept of simultaneity: 
\begin{quotation}
\noindent {\small ``The concept does not exist for the physicist until he has the possibility of discovering whether or not it is fulfilled in an actual case. We thus require a definition of simultaneity such that this definition supplies us with the method by means of which, in the present case, he can decide by experiment whether or not both the lightning strokes occurred simultaneously. As long as this requirement is not satisfied, I allow myself to be deceived as a physicist (and of course the same applies if I am not a physicist), when I imagine that I am able to attach a meaning to the statement of simultaneity. (I would ask the reader not to proceed farther until he is fully convinced on this point.)'' \cite[p. 26]{Einstein20}}
\end{quotation}

A physical concept must be designed in order to bring into unity the multiplicity of experience. In order to do so, there are two main conditions which are essential for the construction of any consistent and coherent physical concept that attempts to represent reality, namely, operational-repeatability and operational-invariance. First, {\it operational-repeatability} points to the fact that a physical concept must be able to bring into unity the multiplicity of physical phenomena observed in different subsequent tests. If every time we observe something it refers to something different, it becomes then impossible to keep track of anything. This is a problem which is well known since Heraclitus' theory of becoming. If there is no repeatability, the reference of different experiences is precluded right from the start and just like in the famous story by Jorge Luis Borges, {\it Funes the Memorious}, the necessary link between the observation of a `the dog at three-fourteen' and `the dog at three fifteen' is completely lost (see for a detailed discussion \cite{deRonde20b}). 
\begin{dfn}
{\sc Operational Repeatability:} A physical concept must be able to bring into unity the multiplicity of physical phenomena observed in different subsequent tests. 
\end{dfn}
Second, {\it operational-Invariance} points to the fact that a physical concept must be also able to provide a consistent ground for the operational testability with respect to different frames of reference (or perspectives). Observers from different perspectives should agree about what they observe. The `dog observed form a profile' should be considered as {\it the same} `dog observed from the front'. In mathematical terms, this means there must exist an operational-invariant formalism which allows us to discuss what is {\it the same} independently of any considered reference frame. 
\begin{dfn}
{\sc Operational Invariance:} A physical concept must be also able to provide a consistent unified account of the operational testability considered with respect to different frames of reference (or bases). 
\end{dfn}
Without these two conditions, allowing us to discuss about experience not only from different instants of time but also from different perspectives, the possibility to produce a coherent and consistent linguistic discourse about reality becomes precluded right from the start. In classical physics, the concepts that secure the tenability of these conditions are the physical concepts of `particle', `wave' and `field' supplemented by the Galilean transformations. We might let Einstein conclude that:
\begin{quotation}
\noindent {\small ``By the aid of speech different individuals can, to a certain extent, compare their
experiences. In this way it is shown that certain sense perceptions of different
individuals correspond to each other, while for other sense perceptions no such
correspondence can be established. We are accustomed to regard as real those
sense perceptions which are common to different individuals, and which
therefore are, in a measure, impersonal. The natural sciences, and in particular,
the most fundamental of them, physics, deal with such sense perceptions.'' \cite[p. 1]{Einstein22}}
\end{quotation}

Unfortunately, in the context of QM, these conditions have been wiped out from the theory. While it has been argued that {\it operational-repeatability} is unattainable due to the fact `quantum particles' are destroyed within each measurement, {\it operational-invariance} has been simply replaced by Bohr's famous {\it principle of complementarity}. On the one hand, the idea that quantum particles are destroyed each time they are measured not only precludes the possibility to consider a repeatable analysis of the object under study, it also gives an alibi to justify why we have no clue of what quantum particles really are. On the other, complementarity has allowed to naturalize not only the inconsistent representations of a state of affairs in terms of `waves' and `particles' ---none of which is linked to the mathematical formalism---, but also to build a new ``contextual common sense'' which precludes the possibility to describe the properties of quantum systems from different frames of reference (or bases). Lacking both operational-repeatability and operational-invariance the orthodox discourse of QM which makes reference to ``quantum particles'' has become nothing more than a bad metaphor, a fictional story with no link to the mathematical formalism nor to the experience we observe in the lab. 



Since the tragic slaughter of operational-repeatability and operational-invariance in QM, the theory has been accused of multiple crimes. Apart from its incapacity to represent ``quantum particles'' convincingly, one of the most famous has been its inability to provide {\it certain} predictions. So they say, QM is intrinsically ``random'', an ``unpredictable'' theory which, due to existence of quantum superpositions, is incapable to predict ---apart from very special cases--- the result of a measurement outcome produced by an ``elementary particle''. This has created an unspeakable aversion against superpositions within a community which silently agrees that these quantum weirdos should be somehow controlled or ---even--- removed from the theory \cite{deRonde18}. All these difficulties arise in the theory given we accept that QM should make reference to single measurement outcomes and quantum particles. On the contrary, if we simply recognize the fact that QM talks about intensive (non-binary) patterns all difficulties suddenly disappear and it becomes quite easy to produce an objective-invariant representation of what the theory talks about. From this standpoint, it becomes completely natural to argue that the Born rule should be understood as making reference to intensive patterns represented by values pertaining to the interval $[0,1]$ ---instead of making reference to single `clicks' with possible values $\{0,1\}$. As a matter of fact, Heisenberg's formalism already provides a consistent link between the mathematical formalism and experience, one which has been repeatedly tested in the lab through the statistical analysis of mean values ---which is what QM actually predicts. Since Heisenberg created matrix mechanics following Mach's {\it observability principle}, QM has provided the operational conditions required to discuss measurements in a consistent manner. If we avoid the temptation to talk about `elementary particles' ---i.e., falling pray of dogmatic metaphysics--- or about single `events' ---i.e., falling pray of naive empiricism--- , the theory of quanta can be naturally restored as an objective theoretical (formal-conceptual) representation which is both operationally-repeatable and operationally-invariant. The opinion that an intensive representation of reality is untenable just because it escapes our classical binary representation in terms of an Actual State of Affairs\footnote{As discussed in detail in \cite{deRondeMassri18a}, an {\it Actual State of Affairs} (ASA) can be defined as a closed system considered in terms of a set of actual (definite valued) properties which can be thought as a map from the set of properties to the $\{0,1\}$. Specifically,  an ASA is a function $\Psi: \mathcal{G}\rightarrow\{0,1\}$ from the set of properties to $\{0,1\}$ satisfying certain compatibility conditions. We say that the property $P\in \mathcal{G}$ is \emph{true} if $\Psi(P)=1$ and  $P\in \mathcal{G}$ is \emph{false} if $\Psi(P)=0$. The evolution of an ASA is formalized by the fact that the morphism $f$  satisfies $\Phi f=\Psi$.} ---common to all classical physics, including relativity--- or our empiricist conviction that observations should be understood as referring to single measurement outcomes ---common to the orthodox understanding of theories---, is a belief supported only by dogma. 



What is needed today, is to begin to accept that QM does not make reference to a binary experience consequence of a microscopic realm constituted by unobservable particles. The Born rule does not need to be understood as making reference to the position of single `particles' and `clicks', it can be naturally read as providing objective intensive information of a state of affairs ---yet to be described. In fact, as we have shown explicitly in \cite{deRondeMassri18a}, if we choose to stay close to the quantum formalism and make reference to the intensive data  computed by the theory (i.e., mean values), objectivity and invariance can be easily restored. Since we already have a consistent mathematical formalism, the problem remains to derive a physical concept which can be quantified in terms of a number pertaining to the closed interval $[0,1]$. In \cite{deRonde17, deRonde18, deRondeMassri18a, deRondeMassri18b, deRondeMassri19b} we have presented a new (non-classical) conceptual architectonic which through the notion of intensive power allows us to provide an ({\it anschaulich}) objective-invariant content to the theory of quanta. According to this representation QM talks about a Potential State of Affairs constituted by powers with definite potentia. 

A {\it quantum power} can be understood in terms of a potential action which ---even though has an expression in actuality--- does not require its actualization in order to exist and interact with other powers. Powers exist in a potential mode of existence ---which is not reducible to a teleological reference to actuality--- just in analogous way as objects exist in an actual mode of existence. It is the widespread confusion between metaphysical actuality ---which makes reference to a categorical representation of existence--- and empirical actuality ---which makes reference to here and now observability--- which has allowed to mix up the notion of object with the observation of its profile; i.e., the obvious fact that the adumbration of an object is not the object itself. An object is a {\it metaphysical machinery} capable of unifying multiple experiences. We never observe an object in its theoretical totality. This can be reached only in a conceptual (or representational) level of understanding. Obviously, observing the side of a table does not imply the existence of the table as a whole (see for a detailed analysis \cite{deRonde20b}). Only through multiple observations in the here and now, it is possible to grasp a (metaphysical) object of experience. Indeed, a physical concept has the goal of bringing into unity distinct experiences, it is built in order to see what is {\it the same} within {\it difference}. The intensive powers that we find in QM are, in this particular respect, not very different from the objects we observe in classical mechanics. Just like it is only through multiple adumbrations of an object that we gain knowledge about its constitution, it is only through multiple observations that we can measure the intensive quantification of a power. Thus, the shift required by QM is twofold. While in the mathematical level we must go from a binary representation to an intensive one, in the conceptual level we must move from the notion of binary actual object to that of intensive potential power. The latter switch also implies a metaphysical shift in the consideration of the mode of existence which must now turn from an Actual State of Affairs to a Potential State of Affairs. 

A {\it Potential State of Affairs} can be understood as the set of powers with potentia given formally represented by an abstract matrix, $\rho$, and conceptually understood as all the experimental possibilities of a lab. A quantum superposition, $\sum c_i \ | \alpha_i \rangle$, or a density operator in a specific basis $\rho_B$, can be then regarded as a specific experimental arrangement in the lab; i.e., a {\it quantum perspective} specified by a basis (or reference frame) in which a particular subset of powers can become actualized \cite{deRondeMassri18b}. The important point here is that once we accept an intensive form of quantification, all quantum powers can be considered as existent beings independently of the observation of their adumbration exposed in a measurement outcome. Powers interact between each other ---in a potential realm--- as described by the mathematical formalism. Consequently, one must distinguish between {\it potential effectuations} related to the process of entanglement and {\it actual effectuations} related to measurement outcomes \cite{deRondeMassri19b}. Powers are not binary, they are intensive in nature. Thus, their understanding requires always a statistical level of analysis which captures the possibility of repeatable measurements right from the start. Unlike the case of elementary particles which cannot be measured repeatedly, an intensive power remains the same independently of its particular observations (operational-repeatability) or the basis chosen to measure it (operational-invariance). And since the potential state of affairs does not change due to measurements, it makes perfect sense to claim that a set of powers are objectively real. All this can be consistently tested in the lab. In order to measure different powers we need to observe their actualizations from different quantum perspectives. Of course, in order to measure the potentia of each power we require many repeated measurements. Due to our global intensive representation the context we chose to measure a power has no influence whatsoever in the results we will obtain. This goes in line with the already mentioned operational conditions presented by the Born rule itself. Our definitions are completely objective, in the sense that the potential state of affairs remains always completely independent of the choice of a specific measurement set up or any observed measurement outcome. Unlike the orthodox contextual understanding of QM, our approach restores the necessary global consistency in order to produce an objective representation of a (potential) state of affairs. 

Against the possibility of a realist understanding of modality, Dennis Dieks argued in \cite[p. 133]{Dieks10}: ``I think it is unclear how a realist interpretation of $p$ as some kind of ontologically objective chance can help our understanding of what is going on in nature. Clearly, such an interpretation cannot change the empirical content and predictive power of the theory that is involved.'' On the contrary, we have shown in our referred works how the conceptual understanding of modality as intensive objective information about powers does provide not only an increase in the explanatory capacity of the theory of quanta, but also a radical change in the way in which its empirical content must be considered and analyzed in the context of experimental testing in the lab. Some explicit examples of the manner in which the conceptual representation has inevitable consequences for the analysis of data have been already provided in the context of quantum computation \cite{deRondeFreytesSergioli19} and quantum entanglement \cite{deRondeMassri19b}. We expect that our research proposal will  provide a new path to continue investigating what really needs to be measured according to the theory of quanta. 


\section*{Acknowledgements} 

This work was partially supported by the following grants: FWO project G.0405.08 and FWO-research community W0.030.06. CONICET RES. 4541-12 (2013-2014).

\end{document}